  \documentclass[preprint,12pt]{elsarticle} 

\usepackage{graphicx} 

\begin{document}

\begin{frontmatter}

\title{Low-dimensionality energy landscapes: magnetic switching mechanisms and rates}


\author{P. B. Visscher}
\ead{visscher@ua.edu}
\author{Ru Zhu}
\ead{rzhu1@crimson.ua.edu}
\address{Department of Physics and MINT Center, University of Alabama}

\begin{abstract}
In this paper we propose a new method for the study and visualization of dynamic processes
in magnetic nanostructures, and for the accurate calculation of rates for such processes.
The method is illustrated for the case of switching of a grain of an exchange-coupled
recording medium, which switches through domain wall nucleation and motion, but is
generalizable to other rate processes such as vortex formation and annihilation.  The method
involves calculating the most probable (lowest energy) switching path and projecting the
motion onto that path.  The motion is conveniently visualized in a two-dimensional projection parameterized by the dipole and quadrupole moments of the grain.  The motion along that path can then be described by a Langevin equation, and its rate can be computed by the classic method of Kramers (1940).  The rate can be evaluated numerically, or in an analytic approximation -- interestingly, the analytic result for domain-wall switching is very similar to that obtained by Brown in 1963 for coherent switching, except for a factor proportional to the domain wall volume.  Thus in addition to its lower coercivity, an exchange-coupled medium has the additional advantage (over a uniform medium) of greater thermal stability, for a fixed energy barrier.

\end{abstract}


\end{frontmatter}


\section{Introduction}
Calculating switching rates of small magnetic elements is a very important problem in magnetic information storage.  In addition to conventional field-switched media, new ideas for storing information in vortex chirality and polarity \cite{vortex} and spin torque switched elements \cite{MRAM} require calculation of switching rates, both with external fields (modeling the writing process) and without (modeling stability).

Several approaches have been explored for this purpose, all ultimately involving the solution of a Fokker-Planck equation for the probability distribution of the system.  The problem was first approached by Brown \cite{brown} in 1963, who computed the switching rate for a coherent single-domain uniaxially symmetric particle.  In this case the system is described by a single variable, the angle $\theta$ of the magnetization from the easy axis, or alternatively the easy-axis component $m_x$ of the magnetic dipole moment.  Once the problem has been reduced to a one-dimensional one described by a stochastic Langevin-like equation of the form
\begin{equation}
\label{lang}
\frac{dm_x}{dt}= f(m_x) + (\frac{dm_x}{dt})_{\rm{random}}
\end{equation}%
where the variance of the random term is a known function of $m_x$, the rate can be determined by the well-established method of Kramers(1940)\cite{kramers}.  Brown's approach can be generalized somewhat \cite{coffey} but is limited to coherently-switching systems.

If we go beyond coherent systems to a system with a spatially-varying magnetization described by a micromagnetic model with N cells, the configuration space of the system is $2N$-dimensional and very hard to visualize.
A  general method for rate calculation, involving the expansion of the energy function in a power series near the saddle point in 2N-dimensional configurational space (a Hessian matrix) was developed by Langer {\it et al} \cite{Langer} and has been applied to magnetic systems\cite{fiedler}.  However, this approach is computationally demanding (because it involves $2N \times 2N$ matrices), somewhat formal (giving little heuristic insight into what physical factors control the rate), and because it involves only information near the saddle point, cannot be exact in the low temperature limit in the sense that the present approach appears to be.

The approach we will take in the present paper is to map our 2N-dimensional system into a 1D system governed by a Langevin equation, and compute the switching rate of that 1D system by the method of Kramers.   We will develop the method in the context of an exchange-coupled or anisotropy-gradient element, which is known\cite{lu} to switch incoherently by a sort of domain wall motion, from the initial state shown in Fig. \ref{init} to a final reversed state (see Fig. \ref{configs} below).  We use an anisotropy variation $K(x) \propto x^2$, which can be shown to give optimal switching in the limit of a long particle\cite{suess}.  We also ignore magnetostatic fields in this calculation.
\begin{figure}[!htb]
\begin{center}
\includegraphics[width=3 in]{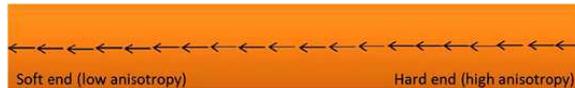} 
\end{center}
\vspace{-0.2 in}
\caption{The 10 nm long anisotropy-graded grain considered in this paper.  The anisotropy field $H_K(x) \propto x^2$, ranging from $0$ at the left ($x=0$) to $9550$ kA/m at the right (very high, but based on numbers for FePt).  We use exchange constant $A = 10^{-11}$ j/m and $M_s=1520$ kA/m.}
\label{init}
\end{figure}%
Although it is very difficult to make direct comparisons with conventional stochastic simulation (because simulation is very difficult for these very slow rates), it seems likely that this method is exact in the limit of low temperatures (large $E_{\rm{barrier}}/k_B T$), because the deviations from the minimum-energy path vanish in this limit.  In this paper we will not deal with the question of global searches for saddle points and switching paths for arbitrary systems -- we are using a relatively simple system (a graded-anisotropy grain) for which the switching mechanism (DW motion) is already known.

\section{Mapping to 1D Langevin motion}
The switching involves motion along or near a one-dimensional path in $2N$-dimensional space, the switching path, parameterized by a single real number, the switching coordinate.  The simplest choice for this switching coordinate is of course $m_x$, the moment along the easy (long) axis of the element, as in the original Brown theory.  In this original theory, $m_x$ determines the energy of the system, since the energy is independent of azimuthal angle.  We could hope that even in an incoherently switching system, this single "coarse-grained" variable still essentially determines the state of the system, in the sense that if we specify $m_x$, fluctuations in the other variables (except the irrelevant azimuthal angle) are small, because they are expensive in energy.  That is, $m_x$ determines values for the other $2N-2$ variables (obtained by minimizing the energy at fixed $m_x$) from which fluctuations are small, because the energy function rises rapidly when these other variables deviate from their minimum-energy values.  This picture turns out to be true near the initial state, as a reversed domain forms, and as the domain wall moves.  However, when this domain wall approaches its highest energy (when it is in the hard end of the system, see inset labeled "saddle point" in Fig. \ref{configs} below) $m_x$ no longer specifies the state of the system uniquely.  There are many states with the same $m_x$, and about the same energy.  One way to see this is to observe that in this configuration, we can increase $m_x$ by moving the domain wall to the right (tilting the magnetizations near the domain wall counterclockwise), {\bf or} by shortening the tail of the domain wall (tilting the magnetizations in the soft end counterclockwise). Near the saddle point, neither of these costs much energy.  Doing these two things at the same time leaves
\begin{equation}
\label{mx}
m_x=\int{M_x dx dy dz}
\end{equation}
unchanged.  Thus a single coarse-grained variable is not sufficient to describe the system -- we need to identify a second variable and hope that the state is determined by the two variables together.  Noting that the two states of increased $m_x$ mentioned above have the increase in different places (in the hard end and the soft end respectively, {\it i. e.} at positive and negative x, if we take our origin at the center of the grain), we see that they will have different values of the weighted average
\begin{equation}
\label{qxx}
q_{xx}=\int{x M_x dx dy dz}
\end{equation}
Note that this is just the quadrupole moment.
Thus we can hope to describe the system by the two coarse-grained variables $m_x$ and $q_{xx}$, in the sense that if we fix these two variables and minimize the energy with respect to the remaining $2N-2$ variables, the fluctuations of these remaining variables (except, again, the azimuthal angle) are small.  We can think of these remaining variables as "short wavelength spin waves".  Denoting this minimum value by $E(m_x,q_{xx})$, we can plot the contours of constant energy as functions of $m_x$ and $q_{xx}$ (Fig. \ref{configs})
\begin{figure*}[!htpb]
\begin{center}
\includegraphics[height=3.5 in]{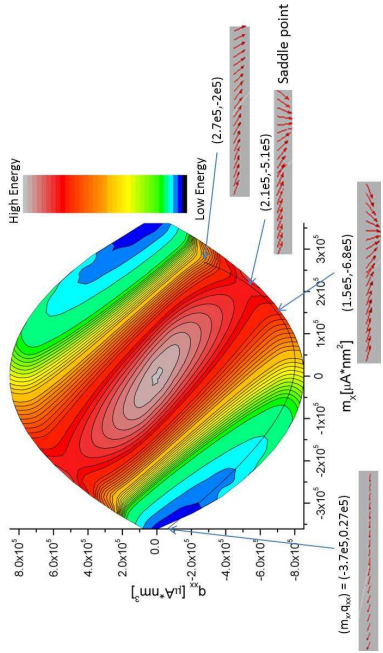} 
\end{center}
\vspace{-0.25 in}
\caption{Contours of the energy function $E(m_x,q_{xx})$, showing also an approximate steepest-descent switching path.  Insets show the actual minimum-energy configuration at various points in the $m_x$-$q_{xx}$ plane.  The boundaries of the colored region (maximum or minimum $q_{xx}$) are known analytically -- they represent configurations with infinitely sharp domain walls (Fig. \ref{sharp}).  Contours are produced from data on a grid, by the Origin \texttrademark graphing package.  They cannot be accurate near the boundaries because the energy diverges in that limit, although they are accurate near the switching path.}
\label{configs}
\end{figure*}%
\begin{figure}[!htbp]
\begin{center}
\includegraphics[width=3.3 in,angle=0]{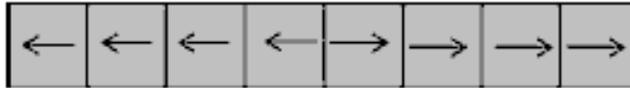}
\end{center}
\vspace{-0.2 in}
\caption{System with infinitely sharp domain wall, represented by the bottom point in Fig. \ref{configs}, with maximum $|q_{xx}|$.}
\label{sharp}
\end{figure}%
In the context of this contour plot, we can understand more clearly the failure of $m_x$ to completely describe the system.  If it did, the energy would rise rapidly as we move away from the minimum in any of the other $2N-1$ directions, including that described by $q_{xx}$.  That is, the energy at fixed $m_x$ would have a sharp minimum as a function of $q_{xx}$.  We can see in Fig. \ref{min} that this is true for the first part of the switching (at the left) but the minima get shallower as we approach the saddle point, finally disappearing entirely at the saddle point.
\begin{figure*}[!htbp]
\begin{center}
\includegraphics[height=3.5 in]{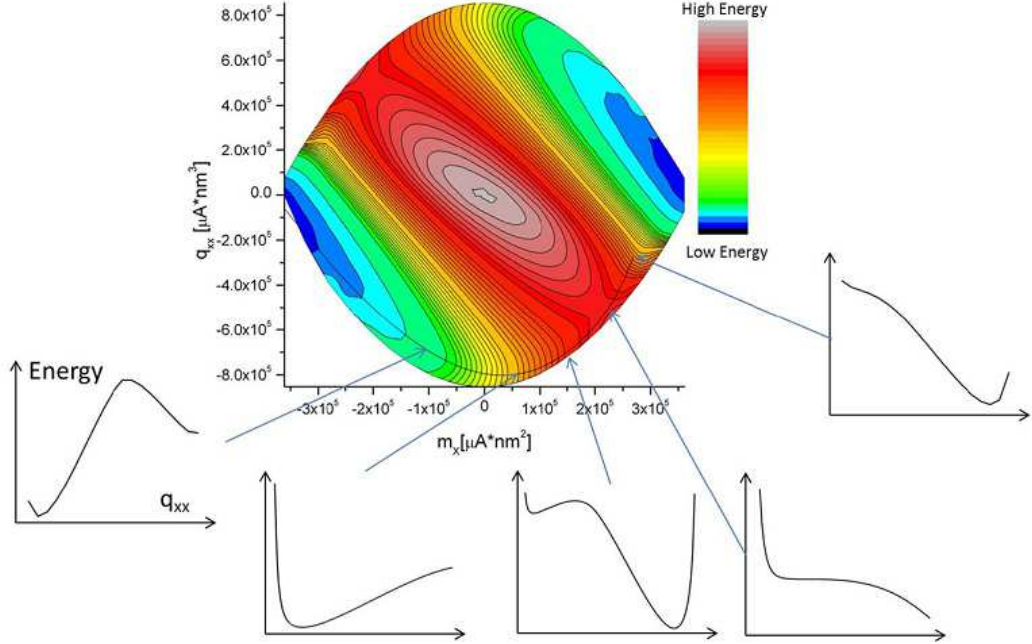}
\end{center}
\vspace{-0.25 in}
\caption{Same energy surface as in Fig. \ref{configs}, with insets showing cross-sections through it at constant $m_x$.  Note that at the beginning of the switching process (on the left) there is a deep energy minimum, meaning that fluctuations about this configuration are small. Near the saddle point, this minimum becomes very shallow and disappears, indicating large fluctuations -- specifying such a value of $m_x$ does {\bf not} determine the state of the system.  (When there are two minima, we are interested in the left-hand one -- the other represents a reversed switching path at the top of the figure, related by $M_x \leftrightarrow -M_x$)}
\label{min}
\end{figure*}%

So far, we have a 2D projection of the motion that allows us to visualize it, but we have not defined a unique switching path -- we have seen that minimizing the energy at fixed $m_x$ is {\bf not} a suitable method for doing so.
Another way to define a switching path is by steepest descent in energy.  At the saddle point, where $2N-1$ of the $2N$ principal curvatures of the energy function are positive, there is a unique direction in which the curvature is downward.  Moving in this direction, and continuing along the energy gradient, defines a unique path which eventually reaches the minimum-energy initial saturated configuration (Fig. \ref{init}).  This appears to be the same path that is obtained by "nudged elastic band" methods \cite{NEB,NEBmag}, but we have computed it by following the energy gradient, which is equivalent to the use of the Landau-Lifshitz-Gilbert dynamical equation for the magnetization $\mathbf{M}$ of a single macrospin:
\begin{eqnarray}
(1+\alpha^2)\frac{d{{\mathbf{M}}}}{dt}= -\gamma \mathbf{M}\times \mathbf{H} - %
\frac{\gamma \alpha}{M_s} \mathbf{M}\times (\mathbf{M}\times \mathbf{H}) \label{LL}
\end{eqnarray}%
in the limit of infinite Landau-Lifshitz damping factor $\alpha$, so it becomes
\begin{eqnarray}
\frac{d{{\mathbf{M}}}}{dt}= %
\frac{\gamma}{\alpha M_s} \mathbf{M}\times (\mathbf{M}\times \mathbf{H}) \label{LLdamp}
\end{eqnarray}%
We must let the gyromagnetic ratio $\gamma \rightarrow \infty$ as well, such that the ratio $\gamma / \alpha$ remains finite.  Physically, this eliminates the precession term and causes the system to move along the energy gradient in 2N-dimensional configuration space.

Although the system cannot be described at all points of the switching process by specifying the single variable $m_x$ and allowing the other variables to relax to their minimum-energy values, this does {\bf not} mean that the switching process is not in some sense one-dimensional.  The energy still goes up if we move in the direction normal to the steepest descent (SD) path (along the SD path is the only direction in which it goes down).  If we define a new "switching coordinate" in the $2N$ dimensional space by by projecting each point perpendicularly onto the SD path, then fluctuations from these projected points are indeed small (the energy goes up rapidly perpendicular to this path).  The system can still be described by a single variable, it just can't be $m_x$ in the sense that we described above.  We can parameterize the path in any way we want -- for example, we could use the path length along the SD path as a switching coordinate.   However, this would be numerically complicated, and we could equally well use any monotonic function of this path length.  Paradoxically, $m_x$ is such a monotonic function, though we just argued that it cannot be used to uniquely or almost-uniquely describe the system!  The resolution of this paradox is that we cannot determine the other variables by minimizing the energy in a hyperplane of constant $m_x$ (a vertical plane in Fig. \ref{configs}) -- there  is no sharp minimum.  However, if we minimize the energy in a hyperplane perpendicular to the SD path, there {\bf is} a sharp minimum that determines all the other variables.

This leads us to a prescription for mapping an arbitrary system trajectory onto a random walk along the SD path.  At each point of the SD path, we calculate the energy gradient (which is the tangent to the SD path) and draw the $2N-1$ dimensional hyperplane normal to this direction.  If the trajectory is close to the SD path, it will be on one and only one of these hyperplanes, and this determines its switching coordinate.  [If it is far away, farther than the radius of curvature of the SD path, it could be on two such hyperplanes -- but this point would have very high energy, so we will assume that this is a rare event and worry about it later.]

To describe the statistics of switching, we now must write an equation of motion for the switching coordinate, which should take the form of a Langevin equation.  Since we are parameterizing the SD path by $m_x$, this is an equation for $dm_x/dt$.  To avoid problems related to the Ito-Stratonovich ambiguity\cite{garcia,scholz}, we find it better to discuss the change $\Delta m_x$ during a finite time increment $\Delta t$ of a micromagnetic simulation.  We must be careful how we define the change $\Delta m_x$ from the actual change in configuration (the magnetization change $\Delta{\bf M}_r$ at cell $r$).  We cannot just substitute this change into Eq. \ref{mx} to get $\Delta m_x$, but we must first project the new configuration back to the SD path, and only then use Eq. \ref{mx}.  Being similarly careful about defining the diffusion term (variance of $\Delta m_x$), we end up with a Langevin equation of the form of Eq. \ref{lang}.  Since the dependence on the external field $H_x$ is linear, we may put it into the form
\begin{equation}
\label{lang2}
\Delta m_x = \mu (m_x )[H_x  - H_{pin} (m_x )]\Delta t + \left( \Delta m_x \right)_{random}
\end{equation}
This expression is exactly derived from the Landau-Lifshitz-Gilbert equation -- however, it is important to realize that it is exact only on the 1D SD path.  To the extent that the real stochastic system deviates from this path, it is only approximate.  The accuracy could be checked by examining the actual switching trajectories in a stochastic simulation, although this would be possible only for unrealistically low
$E_{\rm{barrier}}/k_B T$ (in a real device, this should be at least 40).  As $T \rightarrow 0$, we expect the actual trajectories to become arbitrarily close to the SD path.
We have defined the "field mobility" $\mu (m_x )$ as the coefficient of the external field in the function $f(m_x)$ in Eq. \ref{lang}; it is a sum over the cells, whose most important characteristic is that it has contributions only from the domain wall, {\it i. e.,} the contribution depends on the hard-axis component of the magnetization.  In fact, an approximate expression for the field mobility is
\begin{equation}
\mu = \frac{ \gamma M_s}{\alpha} V_{wall}
\end{equation}
where $V_{wall}$ is the volume of the domain wall (a small fraction of the total volume $V_{total}$).
We have also defined the pinning field (the field required to keep the domain wall stationary)
\begin{equation}
H_{pin}(m_x)  \equiv \frac{f(m_x)}{\mu(m_x )}
\end{equation}
The random term is defined by its variance
\begin{equation}
\left\langle {\left( \Delta m_x \right)_{random}}^2 \right\rangle = 2D(m_x)\Delta t
\end{equation}
which is a function of position $m_x$ along the path -- $D(m_x)$ can be calculated from the magnetization configuration $\{\mathbf{M}_r\}$.
It is related to the mobility by an Einstein relation,
\begin{equation}
kT\mu (m_x ) = \mu _0 D(m_x )
\end{equation}

This 1D motion is slightly more general than Langevin's original formulation, in that the mobility and diffusivity (which are constant in a textbook Langevin equation) are functions of position in our Langevin equation.

We use the method of Kramers\cite{kramers} to compute the rate from our 1D Langevin equation.  This involves beginning with the Fokker-Planck equation for the probability distribution of solutions to Eq. \ref{lang2}.  We assume the distribution near the barrier is in a steady state, in which the probability current
\begin{equation}
\label{j}
J = \rho \mu (H_x  - H_{pin} (m_x )) - D\frac{{d\rho }}{{dm_x }}
\end{equation}
is small and uniform.  Following Kramers, we write the probability density as a function $C(m_x)$ times its equilibrium value:
\begin{equation}
\rho (m_x ) = C(m_x)e^{ - \left[ {E(m_x ) - \mu _0 m_x H} \right]/k_B T}
\end{equation}
so that $C(m_x) \approx 0$ in the switched well ($m_x \approx +m_s$), and is constant in the initial well.  Inserting this into Eq. \ref{j}, letting $H_x=0$, and using the uniformity of the current gives an equation for this constant,
\begin{equation}
C(-m_s) =  \int_{-m_s }^{m_s } {\frac{J}{{D\left( {m_x } \right)}}e^{ + E(m_x )/k_B T} } dm_x
\end{equation}
The switching rate is then $r = J/\textrm{(well population)}$, where in the populated part of the initial well, $C(m_x) \approx C(-m_s)$, so
\[
r = \frac{J}{{\int_{ - m_s }^{+m_s } {\rho(m_x) dm_x } }}%
= \frac{J}{C(-m_s){\int_{ - m_s }^{+m_s } {e^{ - E(m_x )/k_B T} dm_x }} }
\]
\begin{equation}
\label{r}
 r= \frac{1}{{\int_{ - m_s }^{+m_s } {D(m_x)^{-1}%
e^{ + E(m_x )/k_B T} dm_x } \int_{ - m_s }^{ + m_x } {e^{ - E(m_x )/k_B T} dm_x } }}
\end{equation}
The first integral in the denominator is dominated by the vicinity of the barrier, where $E(m_x ) \approx E_\mathrm{barrier}$.  We can bring out the Arrhenius factor to give
\begin{equation}
\label{arr}
r = r_0 e^{-E_\mathrm{barrier}/k_B T}
\end{equation}
where the prefactor ("attempt frequency") is
\begin{equation}
\label{pre}
r_0 = \frac{1}{{\int_{ - m_s }^{+m_s } {D(m_x)^{-1}
e^{ + [E(m_x)-E_\mathrm{barrier}]/k_B T} dm_x } \int_{ - m_s }^{ + m_x } {e^{ - E(m_x )/k_B T} dm_x } }}
\end{equation}

Using Eq. \ref{pre}, we can numerically calculate the integrals appearing in the rate prefactor from our calculations of the states along the SD (steepest-descent) path.  However, it is also interesting to look at the effects of some simple approximations to these integrals, because it sheds light on the connection with the Brown result for coherent switching.  Brown's 1963 result for the prefactor of a uniaxial particle with anisotropy field $H_K$ was\cite{brown}
\begin{equation}
\label{r0}
r_0 = \alpha \gamma H_K \left(\frac{E_b}{\pi k_B T}\right)^{1/2}
\end{equation}
We can put our result in a similar form -- first note that the first integral in the denominator of Eq. \ref{r} is very sharply peaked at the energy barrier, especially at low temperatures.  We can approximate the energy by a quadratic function near its peak
\begin{equation}
\label{quad}
E(m_x)=E_\mathrm{barrier}\left(1-\frac{(m_x-m_x^\mathrm{barrier})^2}{B^2 m_s^2}\right)
\end{equation}
where $B$ is a dimensionless barrier width factor, of order $1$.
Also at reasonably low temperatures, the second integral in the denominator of Eq. \ref{r} depends only on the initial slope of $E(m_x)$ near the initial state, which is the nucleation field $H_\mathrm{nuc}$.

These approximations allow us to evaluate both integrals analytically, giving us
\begin{equation}
\label{r0}
r_0 = \alpha \gamma H_\mathrm{nuc} \left(\frac{E_b}{\pi k_B T}\right)^{1/2} \frac{W}{B}
\end{equation}
Surprisingly, this is identical to Brown's result for the coherent case, except that the nucleation field is playing the role of the anisotropy field, and there are factors of the domain wall fraction $W$ and the barrier width parameter $B$, which should be of order $1$.  The proportionality to the domain wall width is easy to understand physically, because magnetization fluctuations outside the domain wall cannot drive domain wall motion.

\begin{figure}[!htbp]
\begin{center}
\includegraphics[width=6 in]{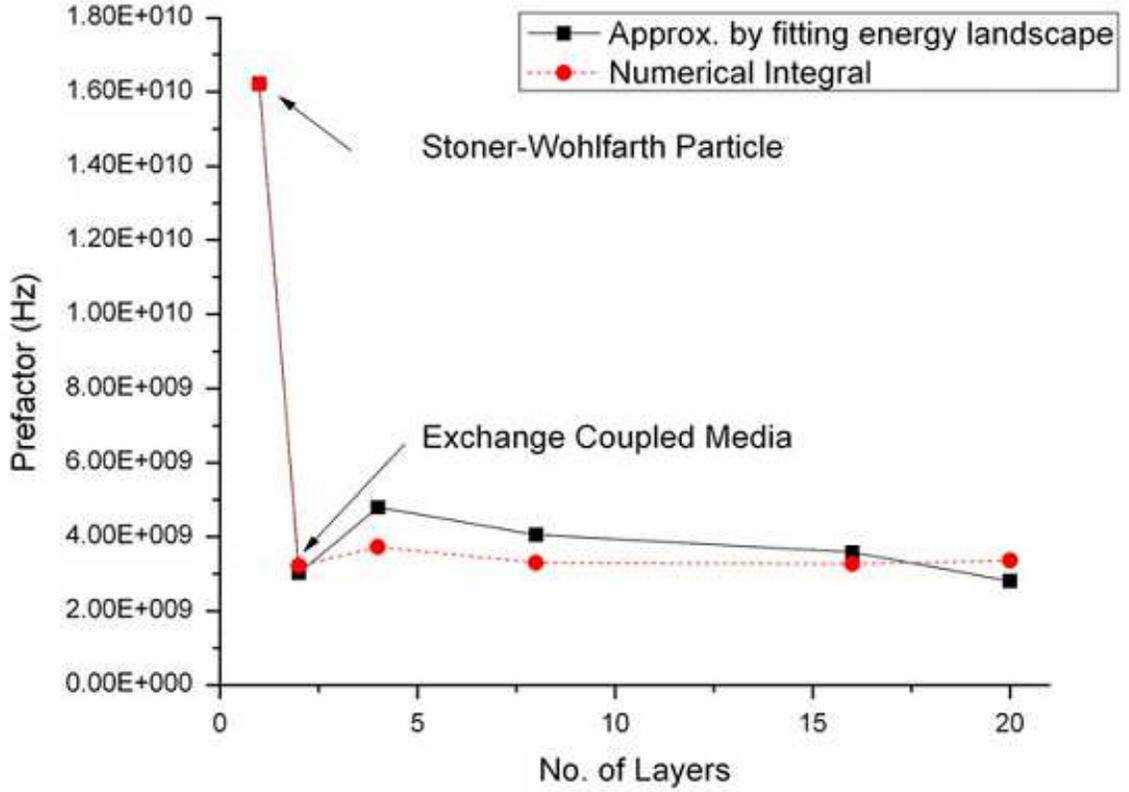}
\end{center}
\vspace{-0.2 in}
\caption{Thermal switching rate prefactor as function of coarseness of discretization (number of layers). The data with black squares are obtained by parabolic fitting of the energy landscape (Eq. \ref{quad}). The red circles are obtained by direct numerical integral of the energy landscape (Eq. \ref{pre}). The switching rate prefactor of the Stoner-Wohlfarth particle (one layer) is obtained by analytical calculation of the energy landscape assuming that the energy barrier is the same as a 20-cell anisotropy-graded system.}
\label{attempt}
\end{figure}

To check the importance of continuous anisotropy grading (as opposed to having layers of uniform but different anisotropy) we considered a series of systems obtained from a continuously graded system by constructing layers and assigning each a uniform anisotropy equal to the average over its volume in the continuously graded system.  As the number of layers increases, we thus expect the behavior to get closer to that of the continuous system.  However, with 1 layer (a uniform system) we expect coherent rotation, then with 2 or more layers the switching mechanism involves a domain wall.  From Eq. \ref{r0}, we expect a large decrease in rate prefactor when a domain wall appears ($W<<1$), and this is confirmed by the numerical results (Fig. \ref{attempt}).  There is suprisingly little change if we increase the number of layers beyond 2.  The advantage of using more than 2 layers lies instead in the improved coercivity.  We have also computed the coercivity as a function of number of layers, and find that the coercivity increases as we decrease the number of layers, because this increases the anisotropy jump at the interfaces and therefore the strength of the domain wall pinning at these interfaces.

\section{Conclusion}
In this paper we have outlined a new method for the calculation of switching rates of incoherently switching micromagnetic systems, and applied it to a simple graded-anisotropy system.  The result can be put into a form that is surprisingly similar to the form derived in 1963 by Brown for coherent switching, except for factors that can be physically interpreted as effects of incoherence, such as the domain wall width.  The predicted rates can be tested by comparison with direct simulation, and such tests are planned.

\section{Acknowledgements}
This work was partially supported by NSF MRSEC Grant DMR-0213985, and by the Defense Advanced Research Projects Agency Office of Microsystems Technology and Grandis, Inc.


\begin{thebibliography}{00}

\bibitem{vortex} S-K. Kim, K-S. Lee, Y-S. Yu, and YY-S. Choi,
Appl. Phys. Lett. {\bf 92}, 022509 (2008).

\bibitem{MRAM}  Z. Li, S. Zhang, Z. Diao, Y. Ding, X. Tang, D. M. Apalkov, Z. Yang, K. Kawabata, and Y.Huai,
"Perpendicular spin torques in magnetic tunnel junctions",
Physical Review Letters {\bf 100}, 246602 (2008).

\bibitem{brown} W. F. Brown, %
"Thermal Fluctuations of a Single-Domain Particle", %
Phys. Rev. {\bf 130}, 1677 (1963).

\bibitem{kramers} H. A. Kramers, %
Physica {\bf VII}, 284 (1940).

\bibitem{coffey} W.T. Coffey, D.S.F. Crothers, J.L. Dormann, L.J. Geoghegan, and E.C.
Kennedy. Effect of an oblique magnetic field on the superparamagnetic
relaxation time. II. influence of the gyromagnetic term. Phys. Rev. {\bf B58} 3249-3266 (1998)].

\bibitem{Langer} J.S. Langer and L.A. Turski,
Physical Review A, {\bf 8} 3230�3243 (1973).

\bibitem{fiedler}
G. Fiedler, J. Fidler, J. Lee, T. Schrefl,
R. L. Stamps, H. B. Braun, D. Suess,
"Direct calculation of the attempt frequency of magnetic structures
using the finite element method",
arXiv:1012.5189v1 [cond-mat.mtrl-sci] (23 Dec 2010)

\bibitem{lu} Z. Lu, P. B. Visscher and W. H. Butler, IEEE Trans. Magn. 43, 2941 (2007).

\bibitem{suess} D. Suess, Appl. Phys. Lett. {\bf 89}, 113105 (2006).

\bibitem{NEB} G. Henkelman, B. P. Uberuaga, and H. Jonsson,
The Journal of Chemical Physics {\bf 113}, 9901 (2000).

\bibitem{NEBmag} R. Dittrich, T. Schrefl, D. Suess, W. Scholz, H. Forster, and J. Fidler,
"A path method for finding energy barriers and minimum energy paths
in complex micromagnetic systems",
J. Magn. Mag. Mat., {\bf 250} (2002).


\bibitem{garcia} J. L. Garcia-Palacios and F. J. Lazaro,
"Langevin-dynamcis study of the dynamical properties of small magnetic particles",
Phys. Rev. {\bf B58} 14937-14958 (1998)].

\bibitem{scholz} W. Scholz, T. Schrefl, and J. Fidler,
"Micromagnetic simulation of thermally acitvated switching in fine particles",
J. Magn. Mag. Mat. {\bf 233}, 296-304 (2001).
\end{thebibliography}
\end{document}